\documentclass[conference]{IEEEtran}
\usepackage[T1]{fontenc}
\usepackage[utf8]{inputenc}
\usepackage{amsmath,amssymb,mathtools}

\usepackage[ruled,vlined,linesnumbered]{algorithm2e}
\SetKwInput{KwInput}{Input}
\SetKwInput{KwOutput}{Output}

\usepackage{xcolor}

\usepackage{dsfont}

\usepackage{cite}
\usepackage{url}
\usepackage{hyperref}
\hypersetup{colorlinks=true,allcolors=black}

\title{High-Performance Reinforcement-Learned BP Decoding of Quantum LDPC Codes
\vspace{-.5cm}}

\author{
Mohsen~Moradi$^*$,
Vahid~Nourozi$^\dag$,
Taejoon~Kim$^*$,
R\'emi~A.~Chou$^\ddagger$,
and David~G.~M.~Mitchell$^\dag$\\
$^*$School of Electrical, Computer and Energy Engineering, Arizona State University, Tempe, AZ 85287, USA\\
(Corresponding author e-mail: mmorad11@asu.edu),\\
$^\dag$Klipsch School of Electrical and Computer Engineering, New Mexico State University, Las Cruces, NM, USA\\ 
$^\ddagger$Department of Computer Science and Engineering, University of Texas at Arlington, Arlington, TX 76019, USA\\ 
\vspace{-.85cm}
}

\begin{document}
\maketitle

\begin{abstract}
Belief-propagation (BP) decoding is attractive for quantum low-density parity-check (QLDPC) codes because it uses local message passing on sparse Tanner graphs. However, conventional flooding BP often stalls due to stabilizer degeneracy and short cycles. Reinforcement-learning-based sequential variable-node scheduling (RL-S), which learns the update order offline, has shown that adaptive scheduling can improve BP convergence. In this paper, we extend this idea with a second-order local update decoder, RL-S2LU. 
The proposed decoder preserves BP locality and low complexity, while numerical results show significant error-correction gains over conventional BP and the considered BP-OSD-10 baseline.
\end{abstract}


\section{Introduction}
Quantum error correction (QEC) protects quantum information by encoding
logical qubits and measuring stabilizer syndromes without directly
measuring the encoded state~\cite{shor1995scheme,gottesman1997stabilizer}.
Quantum low-density parity-check (QLDPC) codes are promising because
their stabilizers are sparse and the number of logical qubits can grow
with the block length~\cite{tillich2014quantum,kovalev2013quantum}.
Recent lifted-product and quantum Tanner constructions provide
asymptotically good families~\cite{panteleev2022asymptotically,leverrier2022quantum},
while bivariate-bicycle (BB) codes demonstrate strong finite-length
performance and low-overhead memory potential~\cite{bravyi2024high}.
These advances make efficient QLDPC decoding a central practical problem.

Belief propagation (BP) is natural for sparse Tanner graphs~\cite{gallager1962low}
and is highly successful for classical LDPC codes~\cite{mackay1999good}.
For QLDPC codes, however, flooding BP often fails because stabilizer
commutation creates many short cycles and degeneracy gives many Pauli
errors with the same syndrome~\cite{poulin2008iterative,panteleev2021degenerate}.
BP with ordered-statistics decoding (BP-OSD) improves reliability by
adding an algebraic post-processing step~\cite{panteleev2021degenerate,roffe2020decoding},
but this moves beyond purely local BP updates.

A complementary approach is to change the BP schedule. Sequential and
shuffled BP reuse newly computed messages within the same iteration and
often converge faster than flooding BP~\cite{zhang2005shuffled,hocevar2004reduced,moradi2026sequential}.
Scheduling can also be learned: reinforcement learning (RL) has been used
to select effective message-passing orders~\cite{habib2023reinforcement,habib2021reldec,moradi2025enhancing}.
For QLDPC codes, RL-S learns a sequential variable-node (VN) update schedule from local residual-syndrome states while leaving the BP update rule unchanged~\cite{moradi2026learning}. 
This approach was subsequently extended to list-based and cluster-based decoding~\cite{moradi2026RL_LS,moradi2026RL_Cluster}.

In this paper, we propose RL-S2LU, a second-order local-update extension
of RL-S. RL-S2LU uses the same trained RL table as RL-S, but changes the
inference-time action: when the policy selects a VN, the decoder updates
that VN and then updates the neighboring VNs that share a stabilizer check
with it. Thus each learned scheduling decision propagates information
farther before the next decision, while remaining local on sparse Tanner
graphs. For the \([[288,12,18]]\) BB code~\cite{bravyi2024high},
RL-S2LU with only \(T=10\) decoder sweeps outperforms BP and BP-OSD-10
with \(T=1000\) iterations, and larger sweep caps further improve the
low-error-rate regime.

\section{Background}
\label{sec:background}

\subsection{CSS QLDPC codes and depolarizing noise}

We consider a CSS QLDPC code defined by two sparse binary matrices
\(
H_X\in\mathbb F_2^{m_X\times n},~
H_Z\in\mathbb F_2^{m_Z\times n},
\)
satisfying \(H_XH_Z^{\top}=0\).  The rows of \(H_X\) and \(H_Z\)
define the \(X\)-type and \(Z\)-type stabilizer checks, respectively.
For a qubit \(v\), let \(\partial_Xv\) and \(\partial_Zv\) be its
neighboring checks in the Tanner graphs of \(H_X\) and \(H_Z\);
for a check \(c\), let \(\partial_Xc\) and \(\partial_Zc\) be the
corresponding neighboring qubits.  Let
\(\mathcal P_1=\{I,X,Y,Z\}\).

Under the independent depolarizing channel, each qubit error
\(E_v\in\mathcal P_1\) satisfies
\(
\Pr(E_v=I)=1-p,~
\Pr(E_v=X)=\Pr(E_v=Y)=\Pr(E_v=Z)=p/3 .
\)
We represent \(E_v\) by its binary \(X\)- and \(Z\)-components
\(
a_v=\mathds{1}[E_v\in\{X,Y\}],\qquad
b_v=\mathds{1}[E_v\in\{Z,Y\}].
\)
Since \(X\)-type checks detect the \(Z\)-component and \(Z\)-type checks
detect the \(X\)-component, the measured Calderbank-Shor-Steane (CSS) syndromes are
\(
s_X=H_Xb \pmod 2,~
s_Z=H_Za \pmod 2 .
\)
The decoder outputs a Pauli estimate \(\hat E\), equivalently binary
components \((\hat a,\hat b)\).  We declare BP convergence when
\(
H_X\hat b=s_X,\qquad H_Z\hat a=s_Z .
\)

\subsection{Sequential BP scheduling}

Flooding BP updates all messages in parallel at each iteration, whereas
sequential BP updates one local part of the Tanner graph at a time and
immediately reuses the new messages.  We use the quaternary sequential
VN update and the residual-syndrome RL state of RL-S~\cite{moradi2026learning}.

\section{RL-S2LU Decoding}
\label{sec:rls2}

RL-S2LU is an inference-time second-order local-update extension of
RL-S~\cite{moradi2026learning}.  It uses the same trained table
\(\mathcal Q_{\rm RL}\), residual state, and greedy
anchor-selection rule as RL-S. 
After selecting a VN, RL-S2LU also updates the VNs that share an adjacent
\(X\)- or \(Z\)-type check with it.

\subsection{Inherited RL-S quaternary VN update}
\label{subsec:rls_qsvns}

We use the two-scalar-stream quaternary BP update of
RL-S~\cite{moradi2026learning}.  The superscript \(X\) labels messages
on the \(H_X\) graph, which estimate the \(Z\)-component \(b\), while
the superscript \(Z\) labels messages on the \(H_Z\) graph, which estimate
the \(X\)-component \(a\).  Both streams are initialized with
\(
\mu=\log\frac{1-2p/3}{2p/3}.
\)
On either Tanner graph \(G\in\{X,Z\}\), the CN-to-VN message is
\begin{equation}
m_{c\to v}^{G}
=
2\tanh^{-1}\!\left(
(-1)^{s_G(c)}
\prod_{u\in\partial_G c\setminus v}
\tanh\!\left(\frac{m_{u\to c}^{G}}{2}\right)
\right).
\label{eq:check_update}
\end{equation}

Let \(a_P\) and \(b_P\) denote the \(X\)- and \(Z\)-components of
\(P\in\mathcal P_1\), and let
\(
p_0=1-\frac{2p}{3},~
\kappa_I=\log(1-p)-2\log p_0,~
\kappa_X=\kappa_Z=\log(1/2)-\log p_0,~
\kappa_Y=\log\frac{3}{4p}.
\)
For scalar beliefs \((L^X,L^Z)\), define the Pauli score
\begin{equation}
\ell_P(L^X,L^Z)
=
\kappa_P+\frac{1}{2}(1-2b_P)L^X+\frac{1}{2}(1-2a_P)L^Z .
\label{eq:pauli_score_compact}
\end{equation}
With \(\operatorname{lse}(x,y)=\log(e^x+e^y)\), the coupled extrinsic
VN-to-CN maps are
\begin{align}
\Phi_X(L^X,L^Z)
&=\operatorname{lse}(\ell_I,\ell_X)
  -\operatorname{lse}(\ell_Y,\ell_Z), \label{eq:phi_x}\\
\Phi_Z(L^X,L^Z)
&=\operatorname{lse}(\ell_I,\ell_Z)
  -\operatorname{lse}(\ell_X,\ell_Y), \label{eq:phi_z}
\end{align}
where all scores are evaluated at \((L^X,L^Z)\).  Thus \(\Phi_X\)
updates the \(H_X\)-stream LLR for \(b\), and \(\Phi_Z\) updates the
\(H_Z\)-stream LLR for \(a\).

A quaternary sequential VN update at \(v\), denoted
\textsc{QSVNSUpdate}\((v)\), first computes~\eqref{eq:check_update} for
all \(c\in\partial_Xv\cup\partial_Zv\), and forms
\[
L_v^{G,{\rm pre}}=\mu+\sum_{c\in\partial_Gv}m_{c\to v}^{G},
\qquad G\in\{X,Z\}.
\]
Then the outgoing messages from \(v\) are updated as
\begin{align}
m_{v\to c}^{X}
&\leftarrow
\Phi_X\!\left(L_{v}^{X,{\rm pre}}-m_{c\to v}^{X},
              L_{v}^{Z,{\rm pre}}\right),
&& c\in\partial_Xv, \label{eq:vn_update_x}\\
m_{v\to c}^{Z}
&\leftarrow
\Phi_Z\!\left(L_{v}^{X,{\rm pre}},
              L_{v}^{Z,{\rm pre}}-m_{c\to v}^{Z}\right),
&& c\in\partial_Zv. \label{eq:vn_update_z}
\end{align}
Afterward, the incoming check messages for \(v\) are recomputed, the
post-update beliefs
\[
L_v^{G}=\mu+\sum_{c\in\partial_Gv}m_{c\to v}^{G},
\qquad G\in\{X,Z\},
\]
are formed, and the hard decision is
\begin{equation}
\hat E_v=\arg\max_{P\in\mathcal P_1}\ell_P(L_v^X,L_v^Z),
\label{eq:pauli_hard_decision}
\end{equation}
with
\[
\hat a_v=\mathds{1}[\hat E_v\in\{X,Y\}],\qquad
\hat b_v=\mathds{1}[\hat E_v\in\{Z,Y\}].
\]

\subsection{Inherited RL-S residual state and anchor rule}
\label{subsec:rls_state}

For a tentative estimate \((\hat a,\hat b)\), define the residual mismatch
vectors
\(
\delta_X=s_X\oplus H_X\hat b,~
\delta_Z=s_Z\oplus H_Z\hat a,
\)
and the total mismatch weight
\(
w=\|\delta_X\|_1+\|\delta_Z\|_1 .
\)
Then \(w=0\) if and only if the estimated Pauli error satisfies both
measured CSS syndromes.

RL-S assigns each candidate VN a local residual state.  Fix a deterministic
ordering of the edges incident to each VN.  Let
\(\operatorname{ord}_X(v,c)\in\{0,\ldots,|\partial_Xv|-1\}\) and
\(\operatorname{ord}_Z(v,c)\in\{0,\ldots,|\partial_Zv|-1\}\) denote the
local positions of \(c\) in \(\partial_Xv\) and \(\partial_Zv\).  Define
\begin{align}
\sigma_X(v)
&=\sum_{c\in\partial_Xv}
  \delta_X(c)\,2^{\operatorname{ord}_X(v,c)},\\
\sigma_Z(v)
&=\sum_{c\in\partial_Zv}
  \delta_Z(c)\,2^{\operatorname{ord}_Z(v,c)}.
\end{align}
Let
\(
A_{\max}=\max\left\{\max_v|\partial_Xv|,\max_v|\partial_Zv|\right\}.
\)
The combined RL state index is
\(
\sigma(v)=\sigma_X(v)+2^{A_{\max}}\sigma_Z(v).
\)
During inference, the trained RL-S table is used greedily:
\begin{equation}
v^\star
=
\arg\max_{v\in\mathcal R}
\mathcal Q_{\rm RL}(\sigma(v),v),
\label{eq:greedy_policy}
\end{equation}
where \(\mathcal R\) is the set of active VNs that have not yet been
selected as anchors in the current sweep.  Let
\(
\mathcal V_{\rm act}=\{v:|\partial_Xv|+|\partial_Zv|>0\}.
\)

\subsection{Second-order local update}

RL-S2LU starts after the greedy rule
in~\eqref{eq:greedy_policy} selects an anchor \(v^\star\).  Instead of
updating only \(v^\star\), RL-S2LU updates \(v^\star\) and its local
second-order VN neighborhood.  For a VN \(v\), define
\begin{equation}
\begin{aligned}
\mathcal U_2(v)
&=
\{u:\exists c\in\partial_Xv \text{ such that } u\in\partial_Xc\}\\
&\quad\cup
\{u:\exists c\in\partial_Zv \text{ such that } u\in\partial_Zc\}.
\end{aligned}
\label{eq:u2_def}
\end{equation}
Thus \(\mathcal U_2(v)\) contains \(v\) and all VNs that share at least
one adjacent \(X\)- or \(Z\)-type check with \(v\).  In the implementation
used for the numerical results, \(\mathcal L_2(v)\) is an ordered
duplicate-free traversal of \(\mathcal U_2(v)\), with \(v\) placed first
and the remaining VNs visited according to the deterministic graph order.

Algorithm~\ref{alg:rls2lu} summarizes the decoder.  The operation
\textsc{ToggleX}\((u)\) is called when the \(Z\)-component \(\hat b_u\)
changes; it flips the residual bits in \(\partial_Xu\), updates \(w\),
updates the affected \(\sigma_X(\cdot)\) values, and marks the affected
remaining VNs for priority refresh.  Similarly, \textsc{ToggleZ}\((u)\)
is called when the \(X\)-component \(\hat a_u\) changes; it performs the
analogous update on the \(H_Z\)-side residuals and states.  

\begin{algorithm}[t]
\footnotesize
\DontPrintSemicolon
\caption{RL-S2LU inference}
\label{alg:rls2lu}
\KwInput{
\(H_X,H_Z\), syndromes \(s_X,s_Z\), physical error rate \(p\),
trained table \(\mathcal Q_{\rm RL}\), sweep cap \(T\).
}
\KwOutput{
\(\hat E\), \((\hat a,\hat b)\), convergence flag, and number of sweeps.
}

Initialize \(m_{v\to c}^{X}=m_{v\to c}^{Z}=\mu\) using $\mu$;
Initialize \(\hat E_v\) for all \(v\) using channel-only beliefs
\(L_v^X=L_v^Z=\mu\) in~\eqref{eq:pauli_hard_decision}\;
Compute \((\delta_X,\delta_Z)\), \(w\), and all states \(\sigma(v)\)\;

\For{\(t=1,\ldots,T\)}{
    \(\mathcal R\gets\mathcal V_{\rm act}\), with key
    \(\mathcal Q_{\rm RL}(\sigma(v),v)\) for each \(v\in\mathcal R\)\;

    \While{\(\mathcal R\neq\emptyset\)}{
        \If{\(w=0\)}{
            \Return{\((\hat E,\hat a,\hat b,\texttt{true},t)\)}\;
        }

        \(v^\star\gets
        \arg\max_{v\in\mathcal R}\mathcal Q_{\rm RL}(\sigma(v),v)\)\;
        \(\mathcal R\gets\mathcal R\setminus\{v^\star\}\)\;

        Form the ordered duplicate-free list
        \(\mathcal L_2(v^\star)\) from \(\mathcal U_2(v^\star)\),
        with \(v^\star\) first\;
        \(\mathcal S\gets\emptyset\)\;

        \ForEach{\(u\in\mathcal L_2(v^\star)\)}{
            \((a_0,b_0)\gets(\hat a_u,\hat b_u)\)\;
            Apply \textsc{QSVNSUpdate}\((u)\) using
            \eqref{eq:check_update}--\eqref{eq:pauli_hard_decision}\;

            \If{\(\hat b_u\neq b_0\)}{
                Apply \textsc{ToggleX}\((u)\), and add the affected VNs in
                \(\mathcal R\) to \(\mathcal S\)\;
            }
            \If{\(\hat a_u\neq a_0\)}{
                Apply \textsc{ToggleZ}\((u)\), and add the affected VNs in
                \(\mathcal R\) to \(\mathcal S\)\;
            }
        }

        Refresh the priority keys
        \(\mathcal Q_{\rm RL}(\sigma(v),v)\) for \(v\in\mathcal S\)\;

        \If{\(w=0\)}{
            \Return{\((\hat E,\hat a,\hat b,\texttt{true},t)\)}\;
        }
    }
}

\Return{\((\hat E,\hat a,\hat b,\texttt{false},T)\)}\;
\end{algorithm}

Only the anchor \(v^\star\) is removed from \(\mathcal R\).  The other
VNs updated inside \(\mathcal U_2(v^\star)\) remain eligible to be
selected as anchors later in the same sweep if they have not already been
selected.  Conversely, a VN that has already served as an anchor can still
be updated later as part of another anchor's local neighborhood.  Hence
\(\mathcal R\) controls only the without-replacement anchor schedule; it
does not mask the deterministic local BP updates.  The LU cascade can
change later residual states and priorities, so RL-S2LU reuses the RL-S
anchor-selection rule and trained table, not necessarily the identical
realized anchor sequence of a separate RL-S run.

\subsection{Complexity and Latency}

We measure complexity and latency in terms of local VN updates. Let
\(
\Delta_v=\max_v\left(|\partial_X v|+|\partial_Z v|\right),~
\Delta_c=\max\left\{\max_c|\partial_X c|,\max_c|\partial_Z c|\right\}.
\)
Then
\[
|\mathcal U_2(v)|\leq 1+\Delta_v(\Delta_c-1)\triangleq U_{\max}.
\]
For bounded-degree QLDPC families, \(U_{\max}=O(1)\). Conventional
flooding BP has \(O(Tn)\) work for sparse Tanner graphs and, under ideal
parallel hardware, \(O(T)\) update latency. RL-S performs sequential VN
updates, giving \(O(Tn)\) local-update work and serial latency.
RL-S2LU performs at most \(U_{\max}\) local VN updates after each selected
VN, so its worst-case work is
\[
O(TnU_{\max})=O(Tn)
\]
for bounded-degree codes, with a larger constant factor than RL-S. Under
a fully serial implementation, the corresponding latency is also
\(O(TnU_{\max})\). However, the updates within \(\mathcal U_2(v^\star)\)
are local and can be parallelized with appropriate synchronization, since
they are triggered by the same selected VN. In such a locally parallel
implementation, the \(U_{\max}\) factor mainly appears as additional
local hardware parallelism rather than as an asymptotic latency factor.
In addition, the sequential selection of \(v^\star\) can be partially
parallelized using clustering or coloring methods.

\section{Numerical Results}
\begin{figure}[!t] 
  \centering
  \includegraphics[width=1\linewidth]{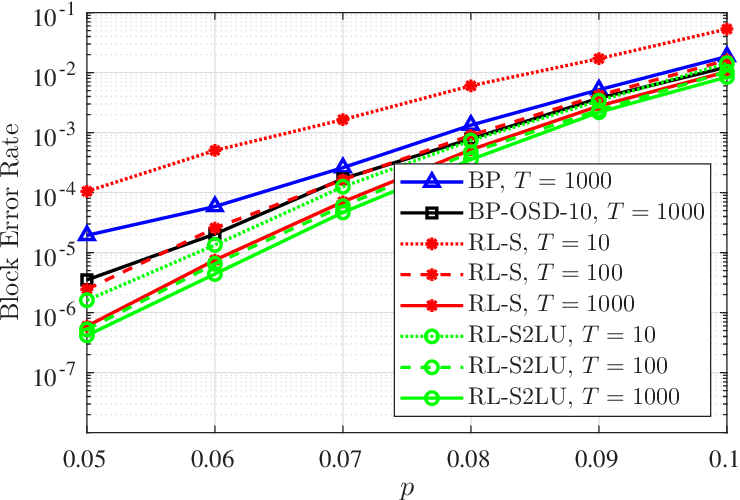}
  \caption{Error-correction performance comparison for the \([[288,12,18]]\) BB code using the proposed RL-S2LU decoder and other state-of-the-art decoders.}
  \label{fig:RL_S2LU_BB288}
\end{figure}

\begin{table}[!t]
\centering
\caption{Percentage of non-convergence errors among the total decoding errors (non-convergence $+$ logical errors) for the \([[288,12,18]]\) BB code corresponding to Fig.~\ref{fig:RL_S2LU_BB288}.}
\label{tab:RL_S2_BB288}
\begin{tabular}{|l||c|c|c|c|c|c|}
\hline
\multicolumn{1}{|c||}{Decoder} & \multicolumn{6}{c|}{\(p\)} \\ \cline{2-7}
\multicolumn{1}{|c||}{} & 0.05 & 0.06 & 0.07 & 0.08 & 0.09 & 0.10  \\ \hline\hline
RL-S2LU, \(T=10\) & 85.2 & 92.5 & 93.8 & 96.0 & 96.7 & 98.0  \\ \hline
RL-S2LU, \(T=100\)  & 41.0 & 73.0 & 85.3 & 90.5 & 93.7 & 97.7  \\ \hline
RL-S2LU, \(T=1000\)  & 27.0 & 60.0 & 67.5 & 85.7 & 89.9 & 92.3  \\ \hline
\end{tabular}
\end{table}

\begin{table}[t]
\centering
\caption{Average number of outer iterations required by the decoding methods considered in Fig.~\ref{fig:RL_S2LU_BB288} for the \( [[288,12,18]] \) BB code under the depolarizing channel.}
\label{tab:avg_iter_BB288}
\begin{tabular}{|l||c|c|c|c|c|c|}
\hline
$p$ & 0.05 & 0.06 & 0.07 & 0.08 & 0.09 & 0.10 \\ \hline\hline
BP, $T=1000$                 & 4.0 & 5.2 & 6.8 & 9.8 & 16.7 & 31.6 \\ \hline
BP-OSD-10, $T=1000$           & 4.0 & 5.2 & 6.9 & 9.9 & 16.6 & 34.2 \\ \hline
RL-S, $T=10$                  & 1.9 & 2.2 & 2.5 & 3.0 & 3.5 & 4.2 \\ \hline
RL-S, $T=100$                 & 1.9 & 2.2 & 2.6 & 3.1 & 4.1 & 6.0 \\ \hline
RL-S, $T=1000$                & 1.9 & 2.2 & 2.6 & 3.6 & 6.7 & 16.0 \\ \hline
RL-S2LU, $T=10$               & 1.0 & 1.0 & 1.0 & 1.0 & 1.0 & 1.2 \\ \hline
RL-S2LU, $T=100$              & 1.0 & 1.0 & 1.0 & 1.1 & 1.3 & 2.1 \\ \hline
RL-S2LU, $T=1000$             & 1.0 & 1.0 & 1.0 & 1.3 & 3.2 & 9.6 \\ \hline
\end{tabular}
\end{table}

We evaluate the decoders using the block error rate, where a decoding failure includes both non-convergence and logical error events; BP-OSD-10 denotes BP followed by order-10 OSD post-processing.
Fig.~\ref{fig:RL_S2LU_BB288} compares RL-S2LU with BP, BP-OSD-10, and RL-S for the \([[288,12,18]]\) BB code of \cite{bravyi2024high} over the depolarizing channel. Even with only \(T=10\) iterations, RL-S2LU outperforms BP-OSD-10 with \(T=1000\) BP iterations over the simulated range. Increasing the iteration cap to \(T=100\) further improves the low-error-rate regime and gives about one order of magnitude gain over BP-OSD-10 at small values of \(p\). Table~\ref{tab:RL_S2_BB288} shows that increasing \(T\) also reduces the fraction of failures caused by non-convergence; for example, at \(p=0.05\), this fraction decreases to \(27\%\) for \(T=1000\). 
Table~\ref{tab:avg_iter_BB288} reports the average number of outer decoding sweeps. For RL-S and RL-S2LU, one outer sweep consists of sequential VN selections; in RL-S2LU, each selected VN \(v^\star\) additionally triggers local updates over \(\mathcal{U}_2(v^\star)\).  Our proposed RL-S2LU typically converges within the first outer sweep in the simulated low-error-rate regime.

\begin{figure}[!t] 
  \centering
  \includegraphics[width=1\linewidth]{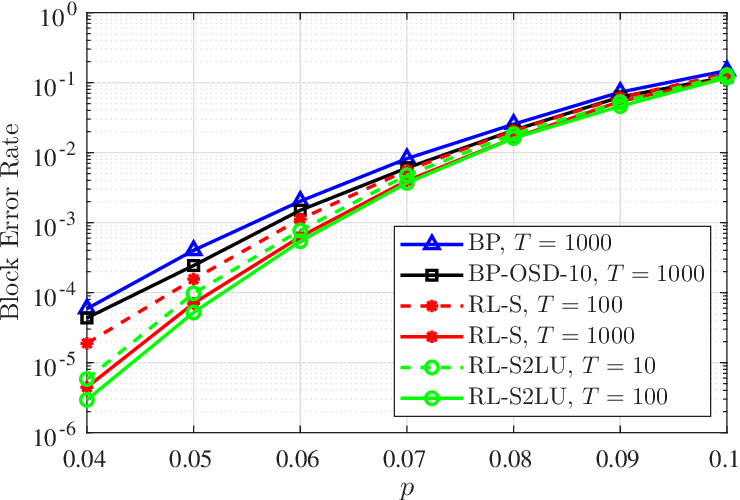}
  \caption{Error-correction performance comparison for the \([[180,10,15\leq d \leq 18]]\) A5 code using the proposed RL-S2LU decoder and other state-of-the-art decoders.}
  \label{fig:RL_S2LU_A5n180}
\end{figure}

Fig.~\ref{fig:RL_S2LU_A5n180} reports the results for the \([[180,10,15\leq d\leq 18]]\) A5 code of \cite{panteleev2021degenerate}. Our proposed RL-S2LU decoder again improves the performance of the underlying RL-S decoder and provides a clear gain over BP and BP-OSD-10. In particular, at \(p=0.04\), RL-S2LU with only \(T=10\) iterations achieves about one order of magnitude lower block error rate than BP-OSD-10 with \(T=1000\) BP iterations.

\begin{figure}[!t] 
  \centering
  \includegraphics[width=1\linewidth]{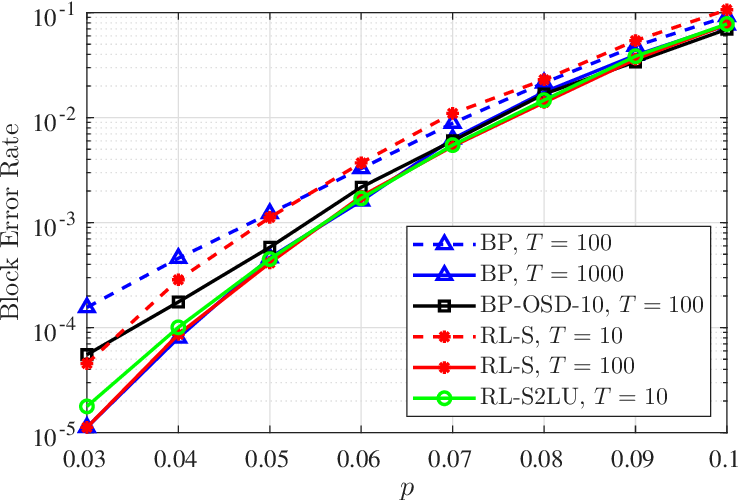}
  \caption{Error-correction performance comparison for the \([[144,12,12]]\) BB code using the proposed RL-S2LU decoder and other state-of-the-art decoders.}
  \label{fig:RL_S2LU_BB144}
\end{figure}

Fig.~\ref{fig:RL_S2LU_BB144} shows the corresponding comparison for the \([[144,12,12]]\) BB code of \cite{bravyi2024high}. For this shorter BB code, RL-S2LU with \(T=10\) already improves over BP-OSD-10 with \(T=100\) at \(p=0.03\), while keeping the decoding process purely BP-based.
For this code, almost all remaining failures of RL-S2LU are logical errors rather than non-convergence errors, suggesting that the block error rate has reached the logical-error-limited regime for the simulated parameters.

\section{Conclusion}
\label{sec:conclusion}
We proposed RL-S2LU, a purely BP-based decoder that replaces each selected
VN action with a deterministic second-order local-update cascade. Numerical
results show improved error-correction performance and faster convergence
on the tested QLDPC codes, indicating that learned sequential BP schedules
can be strengthened by local propagation without retraining or algebraic
post-processing.

\section{Acknowledgment}
This work is in part supported by the National Science Foundation (NSF) under grants CNS2451268, CNS2514415, ITE2515378, and CCF-2145917, and the Office of Naval Research (ONR) under Grant N000142112472.

\bibliographystyle{IEEEtran}
\bibliography{bibliography}

\end{document}